\documentclass[reprint,superscriptaddress,amsmath,amssymb,aip]{revtex4-1}
\usepackage{lineno}\usepackage[normalem]{ulem}
\usepackage{xcolor}\usepackage{tikz}
\usepackage{graphicx}
\usepackage{dcolumn}
\usepackage{bm}
\usepackage[sort&compress]{natbib}
\usepackage{hyperref}

\newcommand{\eps}{\epsilon}

\newcommand{\parf}[2]{\frac{\partial #1}{\partial #2}}

\usepackage{pstricks}

\def\del{\delta}
\def\eps{\varepsilon}\def\hot{{\rm h.o.t.}}
\def\Om{\Omega}\def\pa{\partial}\def\reff#1{(\ref{#1})}
\def\ig{\includegraphics}\def\tew{\textwidth}
\def\huga#1{\begin{gather} #1 \end{gather}}

\newcommand{\btab}[2]{\begin{tabular}{#1}#2\end{tabular}}
\def\hs#1{\hspace{#1}}\def\lam{\lambda}
\begin{document}
\title{Stripes on finite domains: Why the zigzag instability is only a partial story}

\author{Alon Z. Shapira}
 \affiliation{Department of Solar Energy and Environmental Physics, Swiss Institute for Dryland Environmental and Energy Research, Blaustein Institutes for Desert Research, Ben-Gurion University of the Negev, Sede Boqer Campus, Midreshet Ben-Gurion 8499000, Israel}

\author{Hannes Uecker}
 \affiliation{Institute for Mathematics, Carl von Ossietzky University of Oldenburg, P.F 2503, 26111 Oldenburg, Germany}%

\author{Arik Yochelis}\email{yochelis@bgu.ac.il}
 \affiliation{Department of Solar Energy and Environmental Physics, Swiss Institute for Dryland Environmental and Energy Research, Blaustein Institutes for Desert Research, Ben-Gurion University of the Negev, Sede Boqer Campus, Midreshet Ben-Gurion 8499000, Israel}%
 \affiliation{Department of Physics, Ben-Gurion University of the Negev, Be'er Sheva 8410501, Israel}%

\date{\today}

\begin{abstract}
  Stationary periodic patterns are widespread in natural sciences, ranging from nano-scale electrochemical and amphiphilic systems to mesoscale fluid, chemical and biological media and to macro-scale vegetation and cloud patterns. Their formation is usually due to a primary symmetry breaking of a uniform state to stripes, often followed by secondary instabilities to form zigzag and labyrinthine patterns. These secondary instabilities are well studied under idealized conditions of an infinite domain, however, on finite domains, the situation is more subtle since the unstable modes depend also on boundary conditions. Using two prototypical models, the Swift-Hohenberg equation and the forced complex Ginzburg-Landau equation, we consider bounded domains with no flux boundary conditions transversal to the stripes, and reveal a distinct mixed-mode instability that lies in between the classical zigzag and the Eckhaus lines. This explains the stability of stripes in the mildly zigzag unstable regime, and, after crossing the mixed-mode line, the evolution of zigzag stripes in the bulk of the domain and the formation of defects near the boundaries. The results are of particular importance for problems with large time scale separation, such as bulk-heterojunction deformations in organic photovoltaic and vegetation in semi-arid regions, where early temporal transients may play an important role.
\end{abstract}

\maketitle

\textbf{Stationary periodic patterns form in many natural systems, examples of which include electrochemistry, amphiphiles, fluids, chemical reactions, morphogenesis, and vegetation. As such, their formation mechanisms have been studied extensively. However, textbook theory mostly focuses on the analysis on two-dimensional infinite domains, which are an idealization and differ from realistic applications. Using two distinct prototypical models, we show how bounded domains alter, at early time stages, the development of stripes. Specifically, we identify a distinct instability, to which we refer as mixed-mode, and show that stripes can be stable in the mildly zigzag unstable regime, and that deeper in the zigzag unstable regime, it leads to defect formation near the domain boundaries. We believe that the results are significant to applications that involve large time scale separation and where early temporal transients convey important information, such as in organic photovoltaics and vegetation.}

\section{Introduction}
Stationary periodic patterns are abundant in nature and appear at all
scales~\cite{CrossHohenberg1993,maini1997spatial,whitesides2002self,cross2009pattern}. The prototype 
are skin pigmentation in
mammals and fish~\cite{kondo2002reaction, murray2001mathematical}, but periodic patterns appear in
many other systems, ranging from physical and chemical laboratory
setups~\cite{epstein1998introduction,pismen2006patterns}, such as in
nonlinear optics~\cite{arecchi1999pattern}, chemical
reactions~\cite{szalai2012chemical,kapral2012chemical}, and ionic
liquids~\cite{yochelis2015coupling}, to biological and ecological
systems~\cite{murray2001mathematical,meron2015nonlinear}, such as mesenchymal stem
cells~\cite{garfinkel2004pattern} and terrestrial and
underwater
vegetation~\cite{ruiz2017fairy,meron2019vegetation,ruiz2020patterns}. Stationary
periodic patterns with well defined length scales form through a
symmetry breaking that is associated with an instability of a
homogeneous state to 
nonuniform
perturbations~\cite{CrossHohenberg1993,pismen2006patterns,meron2015nonlinear}, 
which, following~\cite{turing1952chemical} is called  `Turing instability' 
or finite wavenumber instability. In two space dimensions (2D), 
the simplest patterns are ``stripes'', which are periodic in one 
direction, say $x$, and homogeneous in the other, say $y$.  
If the stripes bifurcate in the direction
of the unstable uniform state (i.e., as a supercritical bifurcation), 
then the primary stripes with the critical wavenumber will be stable, while 
nearby stripes (with a slightly different wavenumber) will initially 
be unstable but may stabilize at a certain amplitude. 
Conversely, stable stripes may undergo secondary instabilities~\cite{newell1969finite,segel1969distant,busse1978non,greenside1984nonlinear,tuckerman1990bifurcation}, and the
stability region is coined as the ``Busse
Balloon''. 

Secondary instabilities are often also used to explain the evolution to less ordered labyrinthine patterns via stripe bending and/or formation of defects~\cite{pomeau1980wavelength,greenside1984nonlinear,greenside1985stability,yochelis2004two,shiwa2005hydrodynamic,kolokolnikov2006zigzag,kolokolnikov2006stability,burke2007homoclinic,hu2007stability,yochelis2008formation,diez2012instability,uecker2014numerical,lloyd2017continuation,gavish2017spatially}. 
Yet, while infinite domains are useful for analysis, numerical computations are conducted on finite domains, where BCs invoke modes that satisfy only certain symmetries. Moreover, choice of BCs is often physically motivated, and these may show nontrivial implications to the selection of asymptotic (in time) 
patterns~\cite{kramer1984effects,hohenberg1985effects,kramer1985eckhaus,cross1986traveling,chiam2003mean,kozyreff2009influence,dawes2009modulated,doelman2012hopf,rapp2016pattern,verschueren2017model}. In particular, recent applications inspired by electrically charged
self-assembly, indicate that BCs may significantly alter/suppress the development of secondary instabilities of stripes~\cite{gavish2016theory,bier2017bulk}. 
For instance, stability against defects is essential for organic photovoltaic devices, where the loss of efficiency is also attributed to morphological integrity in which stripes break down to segments that preclude transport of electrical charge, see~\cite{shapira2019pattern} and the references therein. Moreover, the {\em transient} evolution of initially
prepared stripes is of paramount significance since the time scale of material evolution is very slow~\cite{reese2010photoinduced,ray2011compact,jorgensen2012stability}.

\begin{figure}[tp]
\btab{l}{(a)\\
\ig[width=0.4\tew]{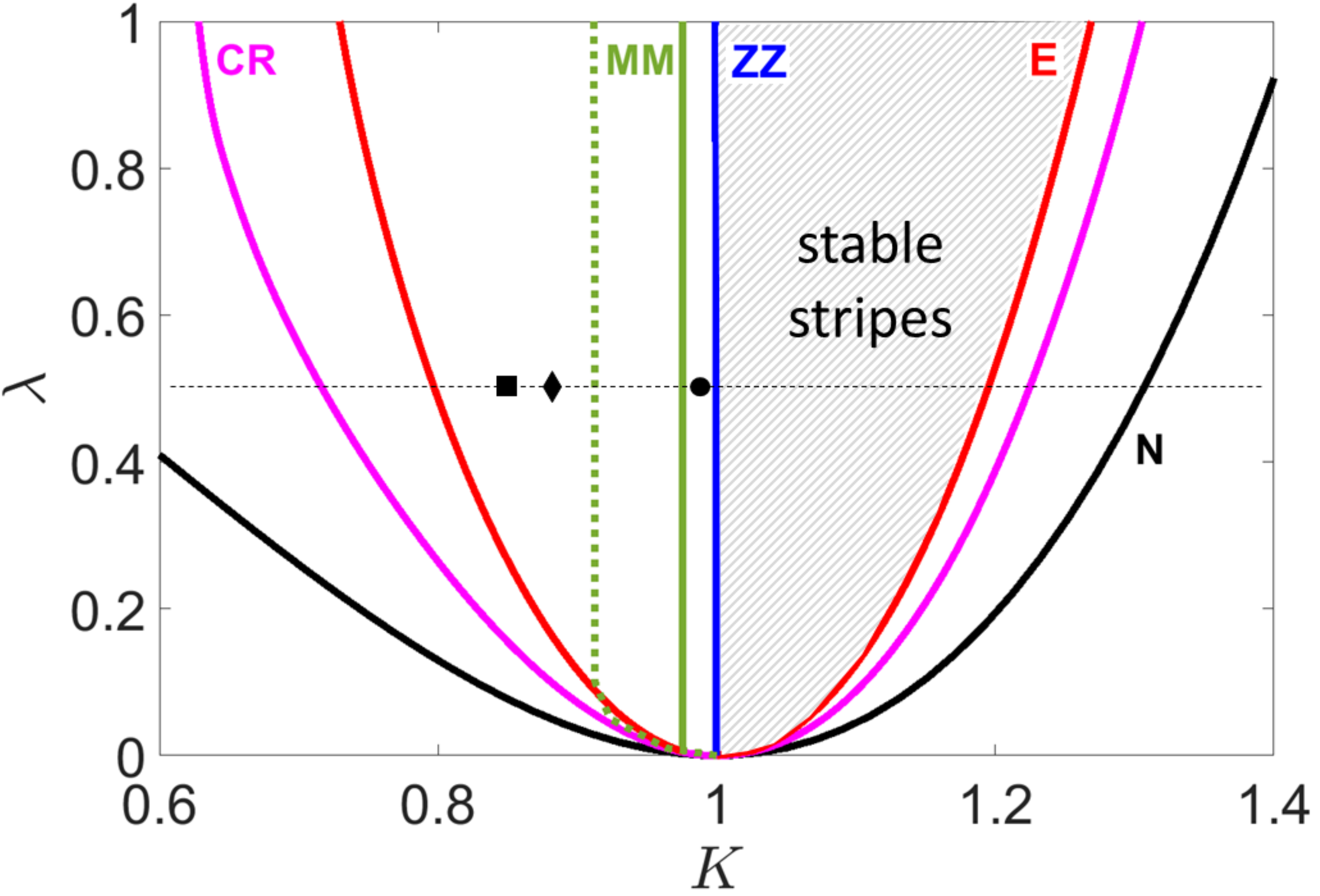}\\
\btab{ll}{(b)&(c)\\
\ig[height=0.14\textheight]{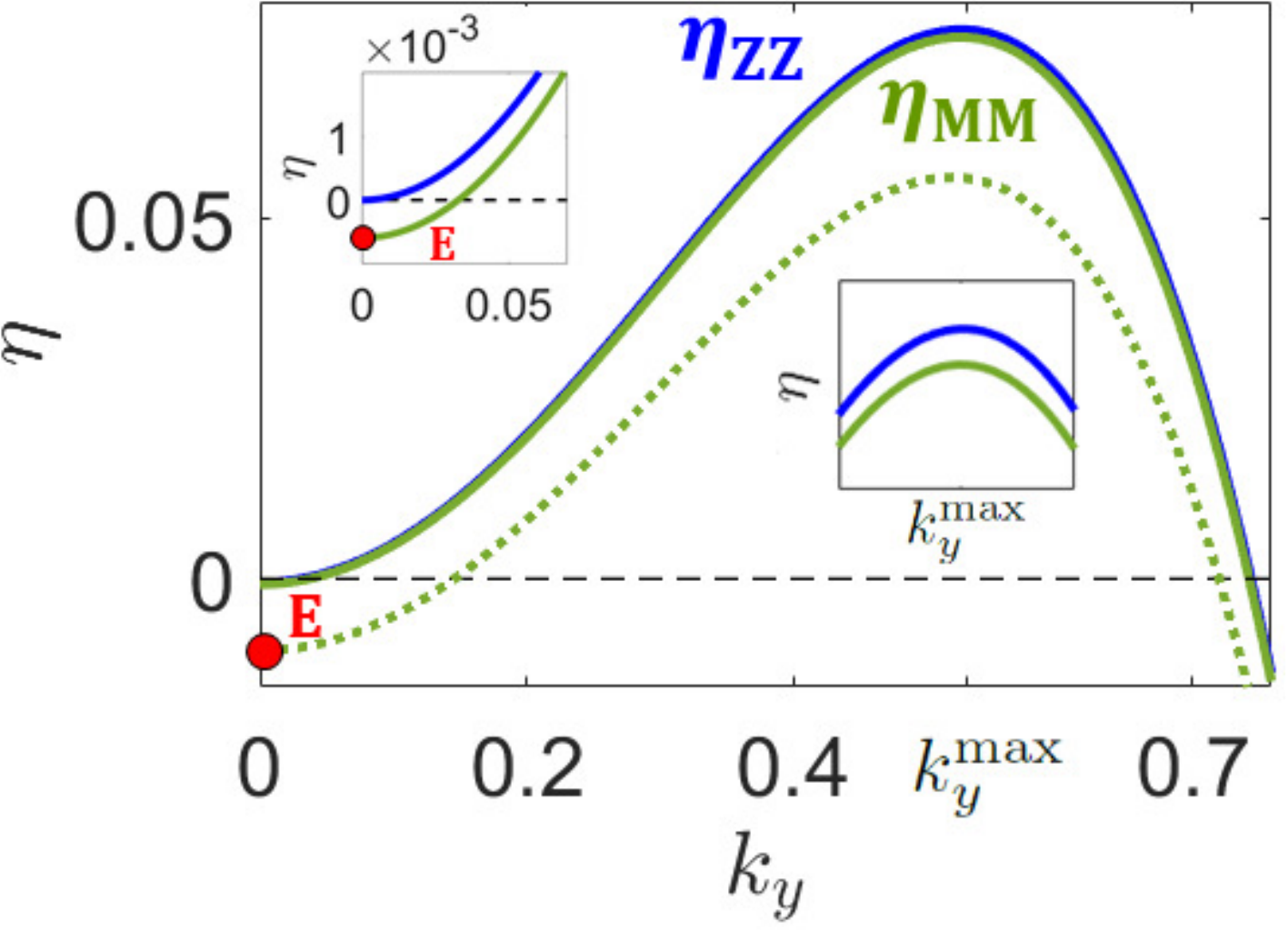}\hs{5mm}
&\ig[height=0.14\textheight]{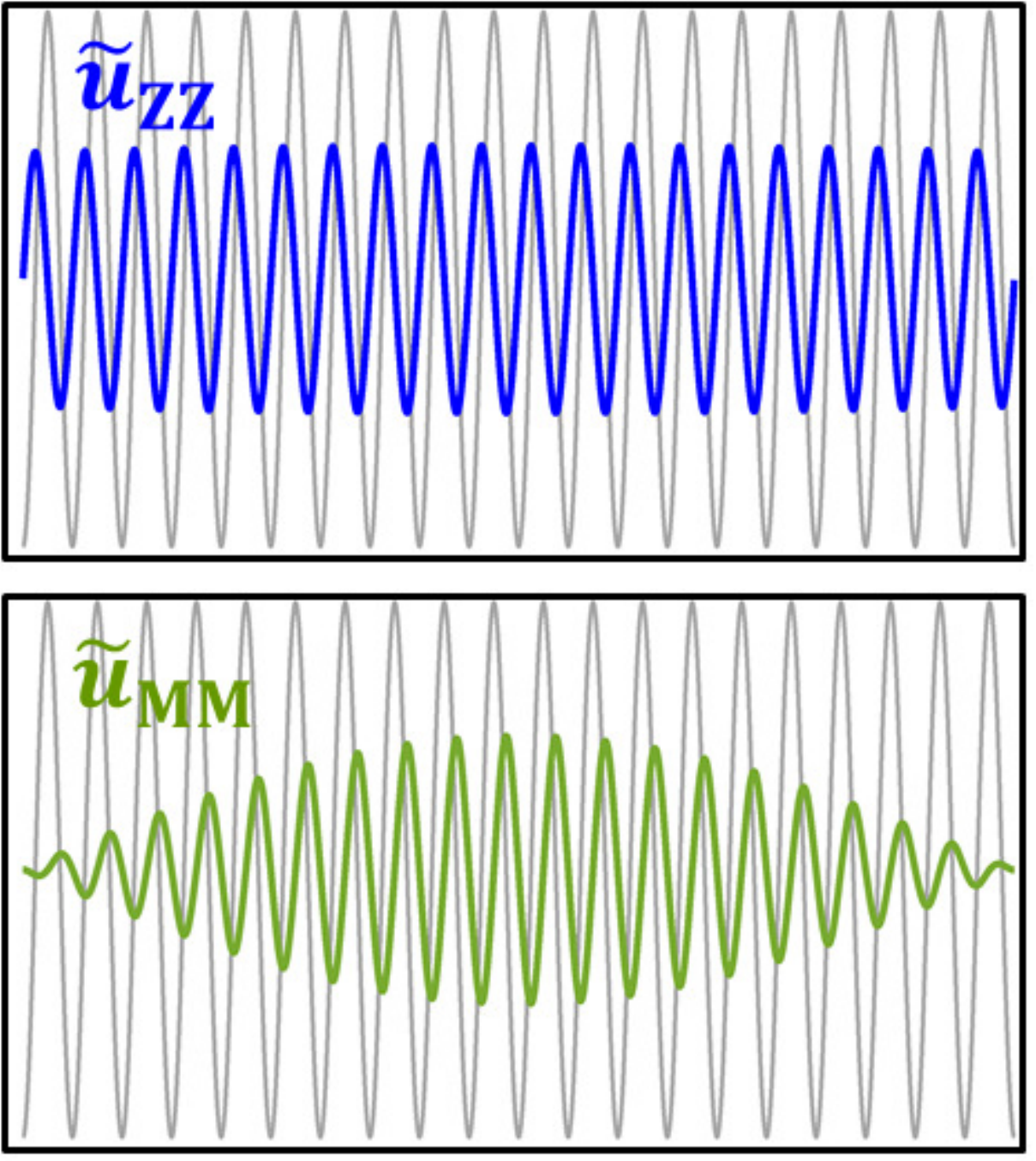}
}\\
\btab{ll}{(d)&(e)\\
\hs{5mm}\ig[height=0.115\textheight]{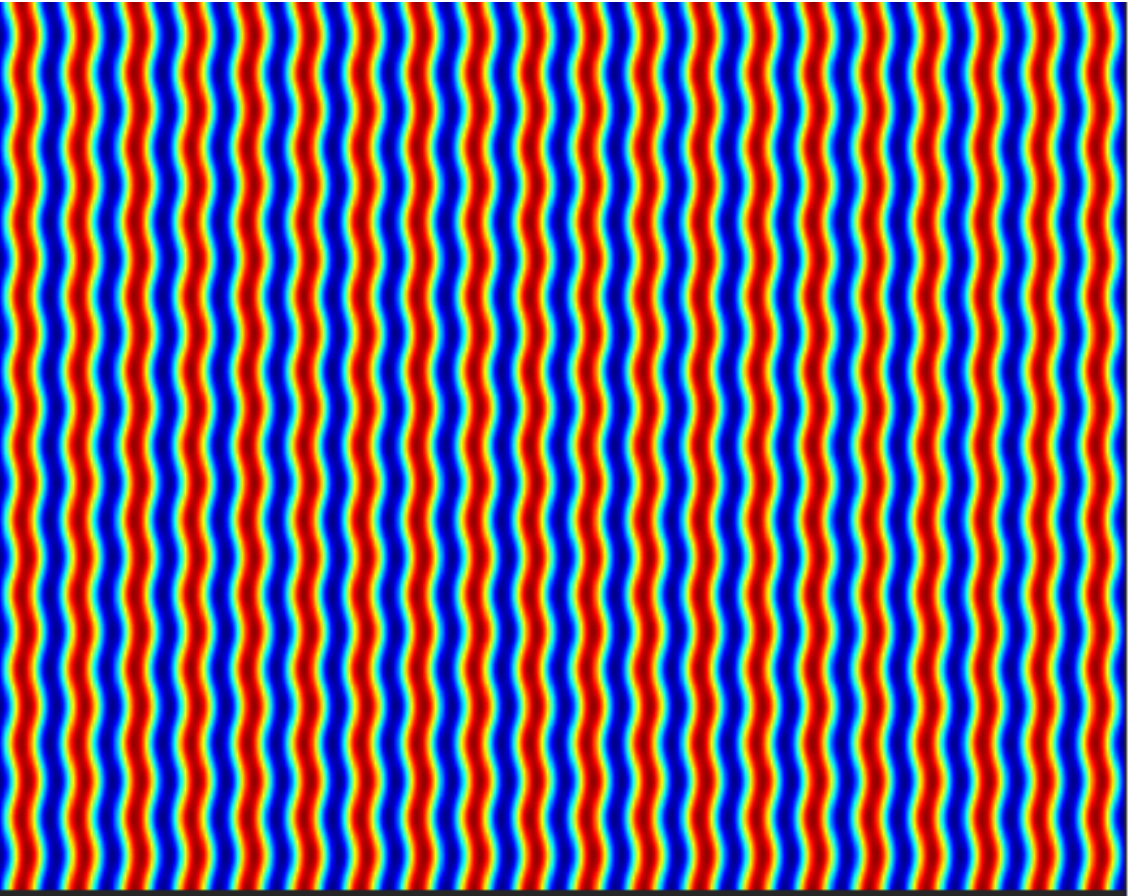}\hs{5mm}
&\ig[height=0.115\textheight]{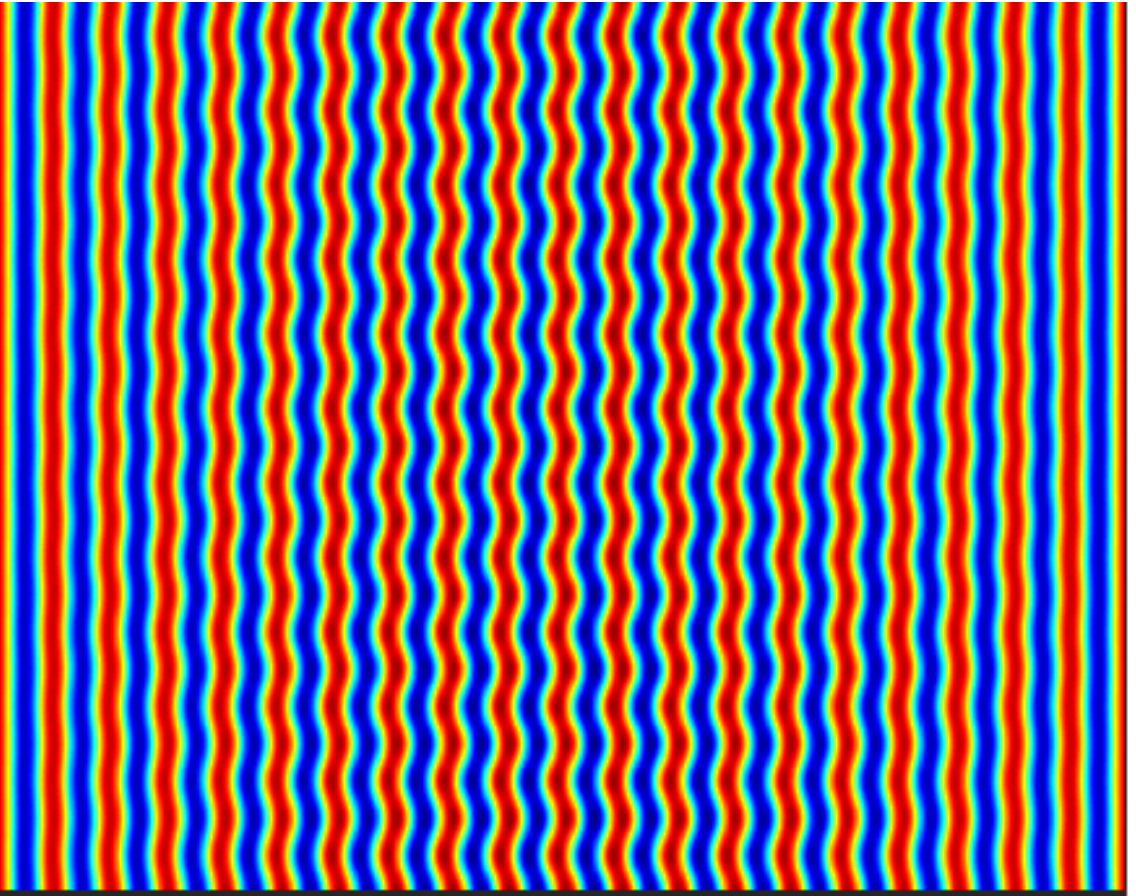}
}}

 \caption{(a) Existence and stability ranges of (periodic) stripe solutions of SH equation~\eqref{eq:she}, where N, ZZ, E, CR, and
    MM stand for the existence, zigzag, Eckhaus, cross-roll, and mixed-mode onsets. The instability onsets have been computed
    numerically via the continuation package \textit{pde2path}~\cite{uecker2014pde2path,dohnalpde2path} and complemented by solving numerically the eigenvalue problem \eqref{eq:evp} with periodic boundary conditions (PBC) for ZZ and otherwise Neumann boundary conditions (NBC). The solid and dashed MM lines (in between E and ZZ for $K<1$) indicate computations on domains consisting of twenty and five periods in $K$. (b) Dispersion relations at $\lambda=0.5$ and $K=0.85$ (`$\blacksquare$' in (a)) computed using~\eqref{eq:evp} for the ZZ ($\eta_\text{ZZ}$) and MM ($\eta_\text{MM}$) instabilities, and the stable Eckhaus mode (E), that is $\eta_\text{MM}(k_y=0)$. The solid line for MM represents computation for $20L_K$ 
and the dashed line is for $5 L_K$ (see also the respective lines in Fig.~\ref{fig:bb}(a)). (c) Respective eigenfunctions $\tilde{u}_\text{ZZ}$ and $\tilde{u}_\text{MM}$ at the maximal growth rate $k_y=k_y^{\max}$, computed on domains with $L_x=20 L_K$ with PBC (top) and NBC (bottom), respectively; the light color periodic solution represents the $u_K$ solution. (d,e) Reconstruction in 2D of the respective ZZ and MM eigenfunctions based on (c).}\label{fig:bb}
\end{figure}

In Fig.~\ref{fig:bb}(a), we show the
textbook diagram with secondary instability onsets for the
Swift-Hohenberg (SH) equation~\cite{swift1977hydrodynamic}
\begin{equation}\label{eq:she}
  \parf{u}{t}=\lambda u-u^3-\left(1+\nabla^2\right)^2u,
\end{equation}
where $u=u(t,x,y)\in \mathbb{R}$, and $\lam$ is an instability parameter. Considering \reff{eq:she} on the infinite 2D domain, `N' is the line above which a family of stripe solutions 
\huga{\label{uk}
u_K(x;\lam)=2\sqrt{(\lam-\kappa^2)/3}\cos(Kx+\phi)+\hot, 
}
exist, with arbitrary phase $\phi$, wavenumber $K$ such that $\kappa=K^2-1\in(-\sqrt{\lam},\sqrt{\lam})$, and where $\hot$ stands for higher order terms. 
Further, `E',`ZZ', and `CR' stand for Eckhaus, zigzag, and cross roll instability onsets, respectively, which 
can be obtained by asymptotic (small $\lam$ and thus small amplitude) analysis~\cite{CrossHohenberg1993,hoyle2006pattern,nepomnyashchy2006general}. Eckhaus instability refers to instability of stripes against parallel stripes (i.e., in $x$ direction) with a slightly different wavenumber $K+\del$, 
where $0<|\del|\ll 1$, i.e., a long wave modulation of the stripe. 

The ZZ instability corresponds to the growth of weak modulations (long wavenumber type) in transverse $y$ direction, while CR is of the finite wavenumber type, associated with the growth of rolls perpendicular to $u_K$. However, on finite domains, unstable modes that do not satisfy the BCs cannot develop, so the picture of secondary instabilities on finite domains requires more subtle treatment.

Here we numerically study, in a paradigmatic setting, two features on
finite 2D domains that are important in applications:  (\textit{i}) A ZZ
instability that may develop under periodic boundary conditions (PBC)
is suppressed under Neumann (no-flux) BC (NBC), and (\textit{ii}) NBC
trigger a distinct secondary instability, 
to which we refer as a \textit{mixed-mode} (MM) instability as it 
combines properties of the ZZ and E modes: The eigenfunction shows 
modulations in $y$ {\em in the bulk of the domain} (ZZ-wise) but the amplitude in $x$ {\em decays towards the boundaries} (E-wise). We employ a numerical linear eigenvalue methodology for spatially extended solutions~\cite{thiele2003front,kolokolnikov2006zigzag,kolokolnikov2006stability,burke2007homoclinic,diez2012instability,gavish2017spatially} to obtain both the {dispersion relations (in $y$ direction) and the respective eigenfunctions (in $x$ direction) that satisfy the basic odd and even symmetries under PBC and NBC, respectively}. We then unfold the link between the eigenfunctions and
the transient evolution from stripes by direct numerical 
simulation (DNS), showing that the most unstable MM 
determines the initial transients, and that the subsequent long 
term evolutions yields defects near the boundary. {This complements 
\cite{greenside1984nonlinear} where the SH equation with 
Dirichlet BC $u=\pa_n u=0$ on all boundaries is studied 
by DNS, where the stripes orient perpendicular to 
the boundaries.} 
For generality, additional to the gradient SH 
model we consider the non-gradient forced complex Ginzburg-Landau (FCGL)
equation and find the same behavior.

\section{The Swift-Hohenberg equation}
The trivial solution $u\equiv 0$ of \reff{eq:she} 
is unstable to waves with wavenumbers $K$ in a band 
around $K_c=1$ such that $(1-K^2)^2<\lam$, and at $\lam=(1-K^2)^2$ 
(the `N' line in Fig. \ref{fig:bb}(a)) there is a supercritical bifurcation 
of stripes of the form \reff{uk} with wavenumber $K$. 
In the following we consider \reff{eq:she} 
on a domain $\Om=(0,L_x)\times (0,L_y)$, with NBC in $x$, $\pa_x u|_{x=0}=\pa_xu|_{x=L_x}=\pa_x^3 u|_{x=0}=\pa_x^3u|_{x=L_x}=0$, 
or PBC 
$\pa_x^ju|_{x=0}=\pa_x^ju|_{x=L_x}, \ j=0,1,2,3$,
which also imply, from \eqref{eq:she}, $\pa_x^j u|_{x=0}=\pa_x^j u|_{x=L_x}$ 
for higher derivatives $j>3$. In $y$, we always use PBC  
$\pa_y^ju|_{y=0}=\pa_y^ju|_{y=L_y}, \ j=0,1,2,3$.
The finite domain with the stated BCs 
has the immediate consequence that only a discrete set of wavenumbers $K$ is admissible (and similar for the wavenumbers $k_y$), but we choose the domains large enough such that 
this discreteness has a minor effect, and which we thus ignore in plots such as Fig.~\ref{fig:bb}(a). 

\begin{figure}[tp]
    (a)~\includegraphics[width=0.175\textwidth]{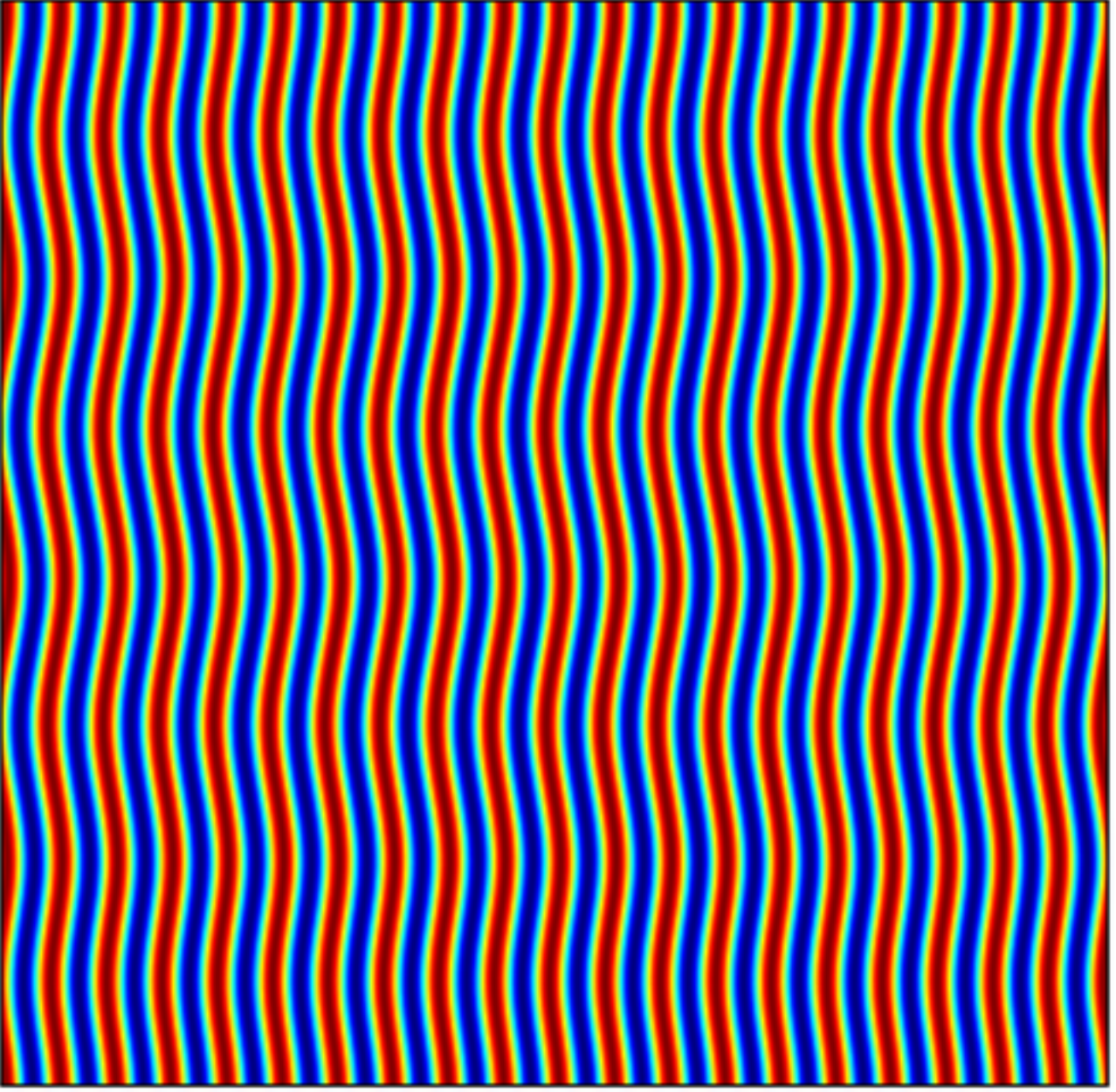} \hskip 0.1in
    (b)~\includegraphics[width=0.175\textwidth]{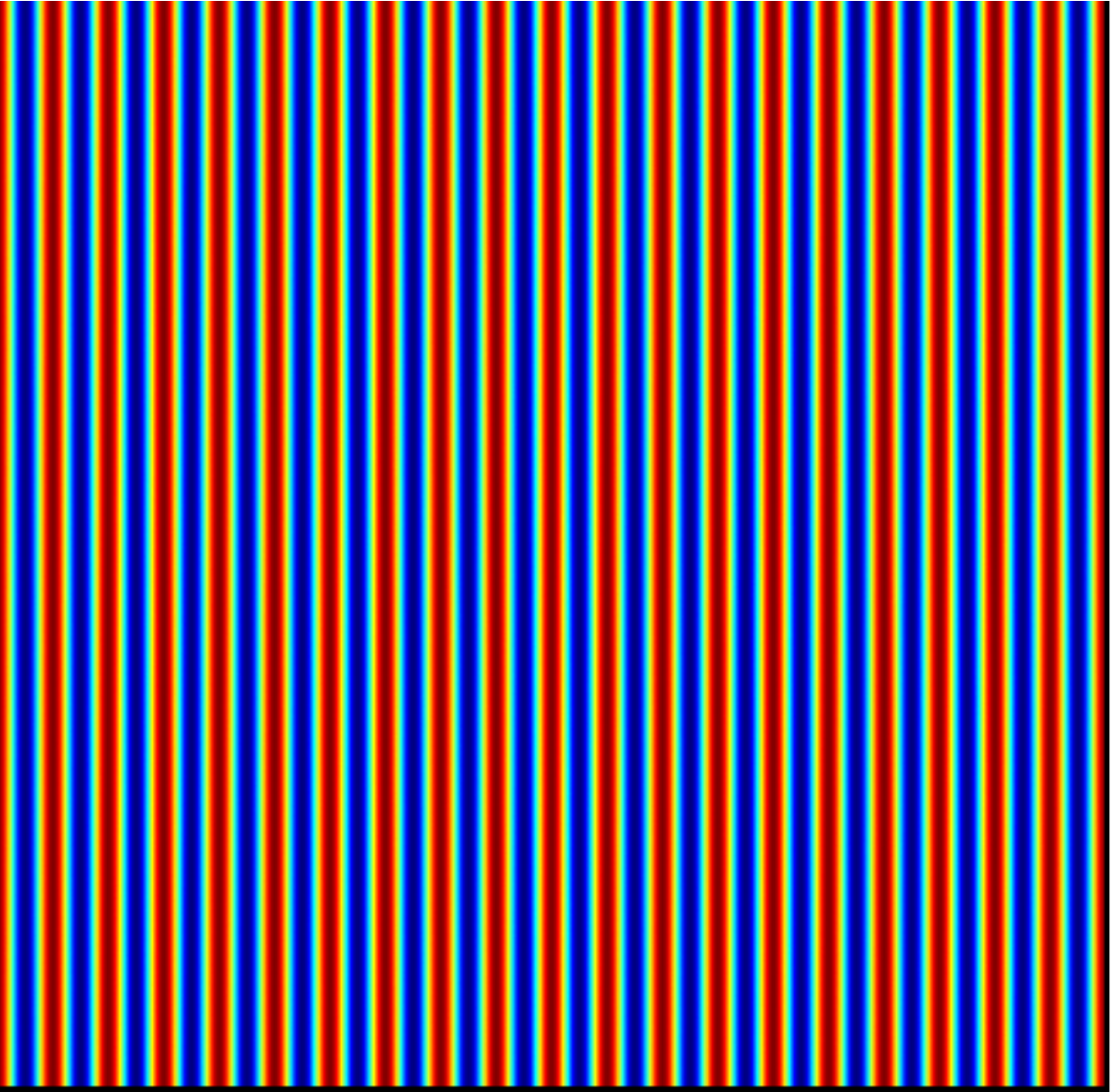}
    \caption{Asymptotic states of stripes obtained by DNS of~\eqref{eq:she} at $\lambda=0.5$ after a random perturbation of $u_K$ with $K=0.98$ close to the ZZ onset (see `$\bullet$' in Fig.~\ref{fig:bb}(a)), with PBC in $y$ direction, and PBC (in (a)) and NBC (in (b)) in the $x$ direction. The domains are
    $\Omega=[0,20 L_K]\times[0,4L_y]$, where $L_K=2\pi/K$,
    $L_y=2\pi/k_y$ and $k_y=0.2$. Colorscale in all plots ranges between $u=-\sqrt{\lam}$ (blue) and $u=\sqrt{\lam}$ (red).}\label{fig:zz_stable}
\end{figure}

The stability of $u_K(x)$ 
is obtained via decomposition in the finite $x$ direction, $\tilde{u}(x)$, and the transverse {infinite} periodic $y$-direction,
with 
wavenumber $k_y$:
\begin{equation}\label{eq:pert}
  u(t,x,y)=u_K(x)+\eps \tilde{u}(x)e^{\eta t+ik_yy}+c.c.+\hot,
\end{equation}
where $\eta$ is the perturbation growth rate, $|\eps|\ll1$ is an auxiliary perturbation parameter, and $c.c.$ stands for complex conjugate. Linearization about $u_K$ results in the eigenvalue problem
\begin{equation}\label{eq:evp}
  \eta\tilde{u}=\left[ \lambda-3u_K^2-\left(1+\partial^2_x-k_y^2\right)^2 \right]\tilde{u}.
\end{equation}
The onsets of the secondary instabilities zigzag (ZZ), Eckhaus (E), and
cross-roll (CR) in Fig.~\ref{fig:bb} 
are obtained numerically for domain length $L_x=20 L_K$ in $x$, 
where $L_K\equiv 2\pi/K$ with PBC in $x$ for ZZ, and
NBC in $x$ for E and CR. The onsets agree well with their analytical expressions, 
e.g., E$(\kappa)=3\kappa^2-\kappa^3+O(\kappa^4)$, $\kappa=K^2-1$ as in \reff{uk}, 
and 
ZZ$(\lam)=-\lam^2/512+O(\lam^3)$, see \cite{am96a}, or~\cite{hoyle2006pattern,nepomnyashchy2006general}. 

The ZZ instability corresponds to the eigenfunction 
$\tilde{u}(x)=u_K'(x)$ in \reff{eq:pert}. Since this violates the NBCs, 
the ZZ instability is replaced by the MM instability that 
in contrast to ZZ, is of finite wavenumber type and is associated with a distinct eigenfunction, as shown in Fig.~\ref{fig:bb}. 
The MM instability onset lies in between the E and ZZ onsets (see Fig.~\ref{fig:bb}(a)) and inherits characteristics of both the ZZ and E instabilities, 
namely, a wavenumber $k_y$ modulation in $y$, which is 
very close to the transverse modulation of the ZZ instability 
(Fig.~\ref{fig:bb}(b)), and the Eckhaus eigenfunction in the $x$ direction, 
which decays towards the boundaries $x{=}0$ and $x{=}L_x$, 
(Fig.~\ref{fig:bb}(c)). Moreover, $k_y=0$ in MM corresponds directly 
to the Eckhaus case (see red dots in Fig.~\ref{fig:bb}(b)) 
so that {\em only} at the E onset $\eta_{MM}(0)=0$. 
Otherwise, $\eta_{MM}(0)$ increases as $L_x\to\infty$, 
but $\eta_{MM}(0)<0$ for all $L_x$, making the qualitative difference 
and justifies to call the MM instability a finite wavenumber instability.
Reconstruction of the ZZ and MM eigenfunctions in 2D via 
\eqref{eq:pert_eig}, illustrates the inherent decay towards the boundaries that is a signature of the E mode 
(Figs.~\ref{fig:bb}(d,e)). The location of the MM instability line in Fig.~\ref{fig:bb}(a) naturally depends on the domain size; for small $L_x$ (dashed green line) it is deep in the ZZ unstable range, while for large $L_x$ (full green line) it is close to ZZ line, and relatedly the MM dispersion relation approximates the ZZ dispersion relation for large $L_x$, see Fig.~\ref{fig:bb}(b). Nevertheless, even on an infinite domain $\eta_{MM}(0)$ is still the Eckhaus mode, and hence $\eta_{MM}$ and $\eta_{ZZ}$ are not identical; they only coincide for $\eta(k_y^{\max})$ in the unstable region. For these reasons, and due to 
the consequences for time evolution discussed next, we prefer the name 
MM rather than 'modified ZZ' or 'modified E' modes.

\begin{figure}[tp]
    (a)~\includegraphics[height=0.145\textheight]{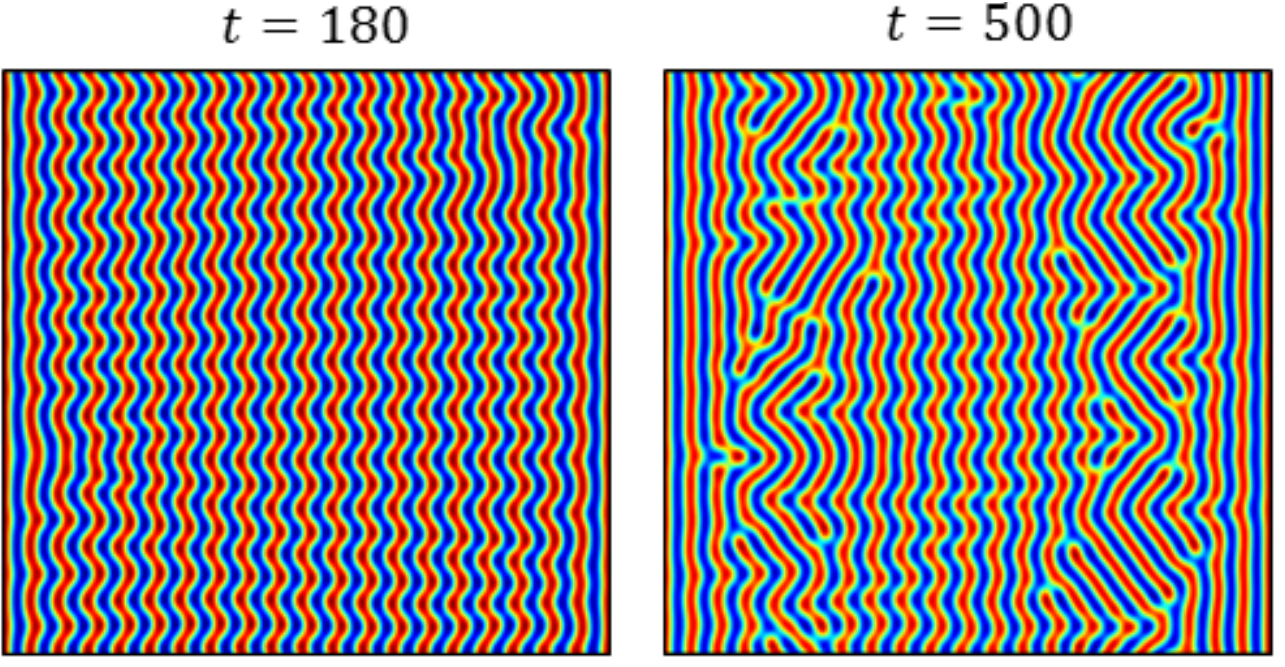} \vskip 0.1in
    (b)~\includegraphics[height=0.145\textheight]{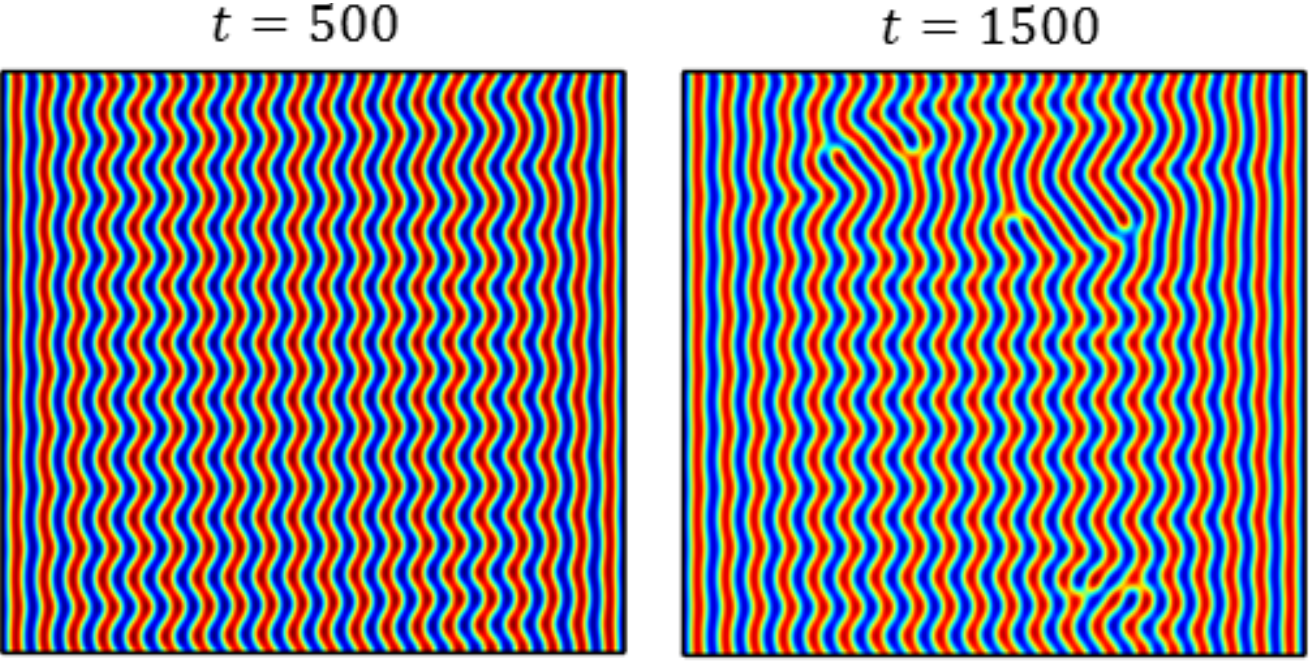}
\caption{Snapshots of DNS  of~\eqref{eq:she} under NBC in $x$ direction, in the MM unstable regime, (a) $K=0.85$ and (b) $K=0.88$, see `$\blacksquare$' 
and `$\blacklozenge$' in Fig.~\ref{fig:bb}(a), respectively; we emphasize that these are not asymptotic solutions. 
The domains are $\Omega=[0,20 L_K]\times[0,12 L_y]$, $L_y=2\pi/k_y$ with 
$k_y=0.53$ in (a), and $\Omega=[0,20 L_K]\times[0,10 L_y]$, 
$k_y=0.47$ in (b).} \label{dnsf1}
\end{figure}

\begin{figure*}[tp]
\btab{l}{(a)\\
\ig[height=29mm]{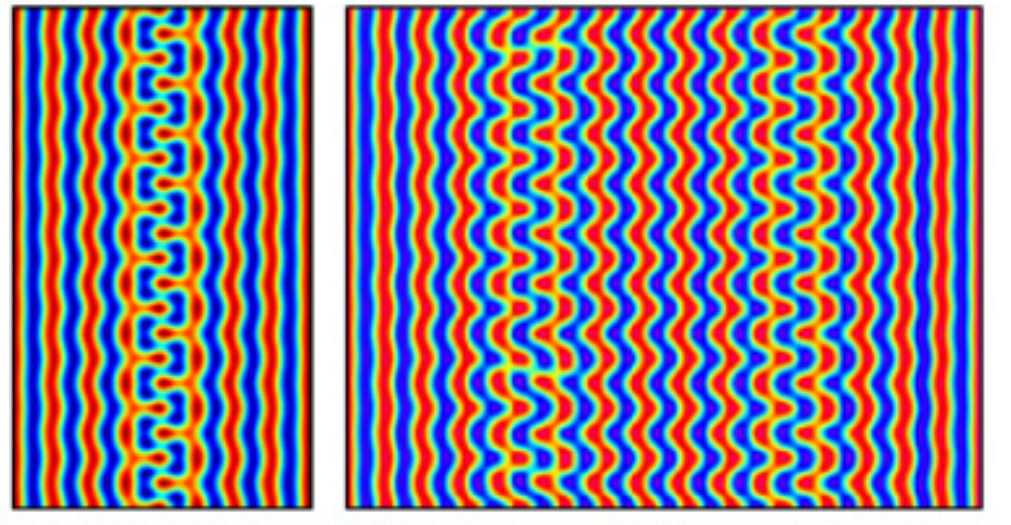} \ig[height=29mm]{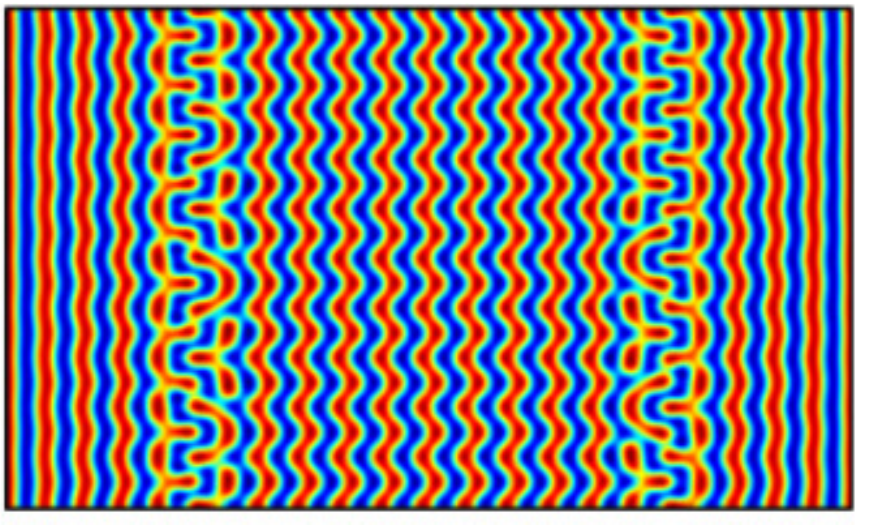}\ig[height=29mm]{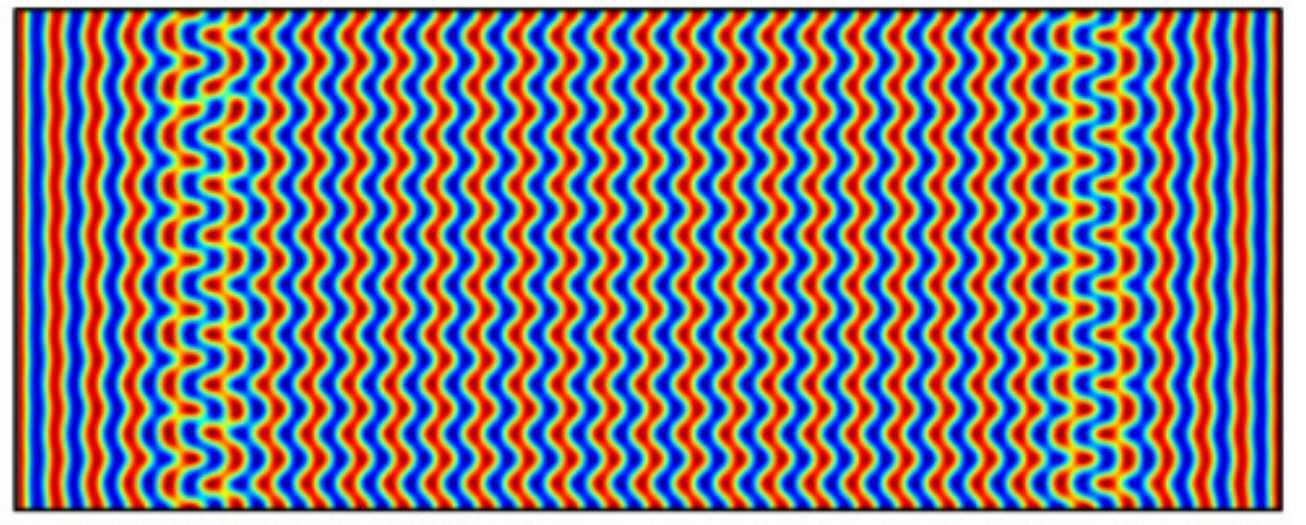}\\
(b)\hs{60mm}(c)\\
\ig[height=40mm]{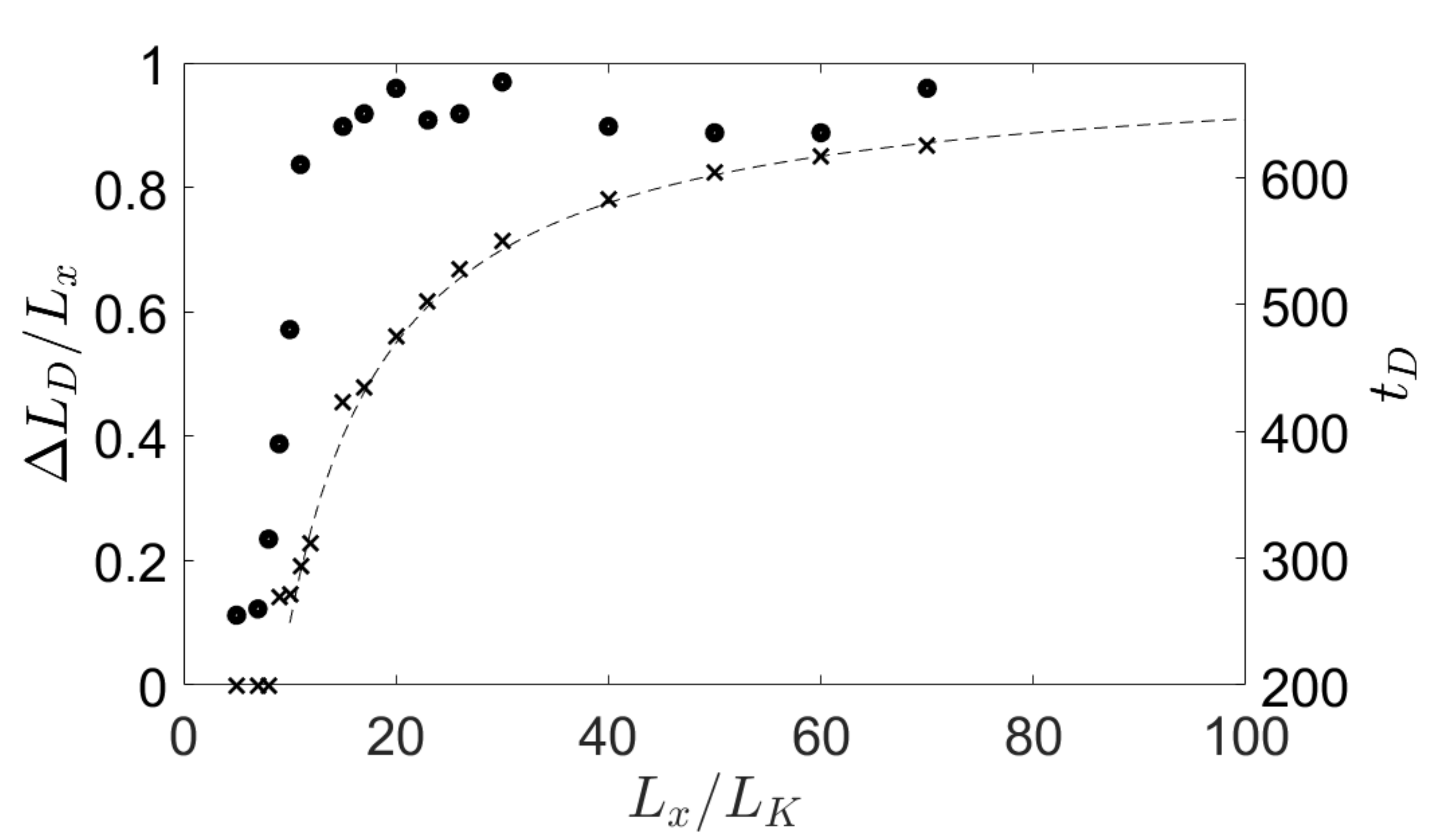}\hs{3mm}
\ig[width=0.6\tew]{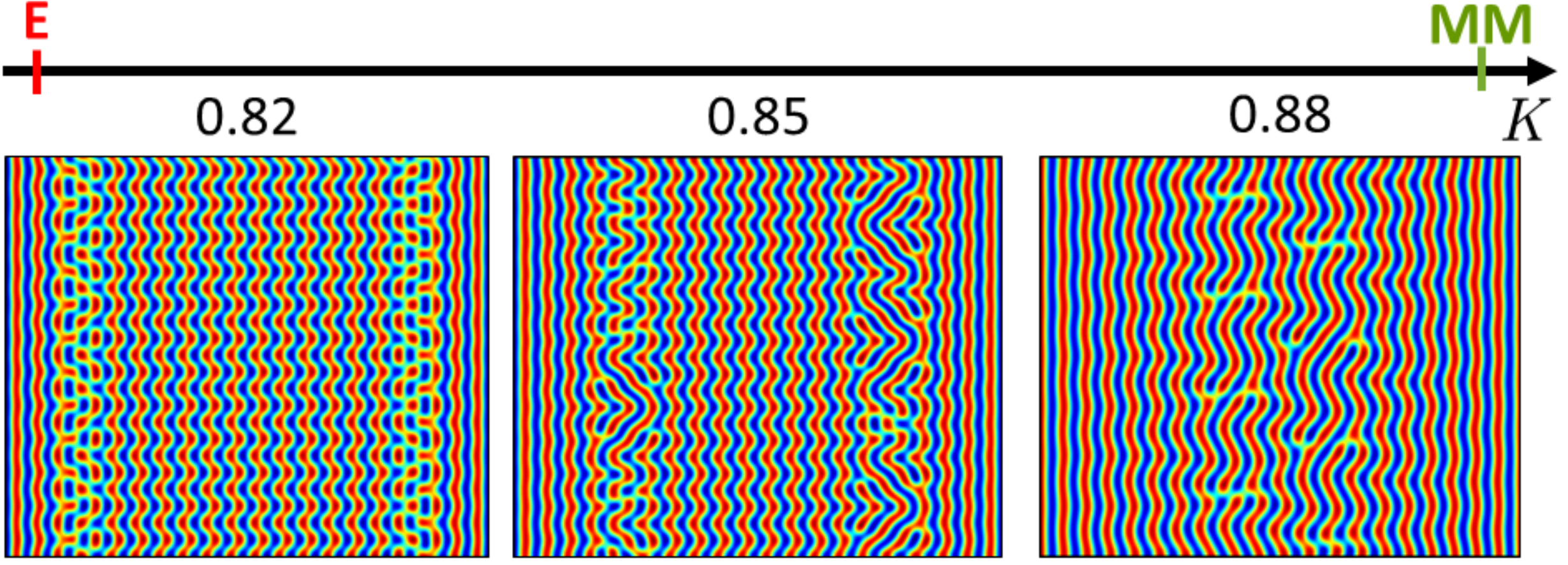}
}
\caption{(a) Snapshots of DNS of~\eqref{eq:she} 
 with~\eqref{eq:pert_eig} as
    initial conditions, NBC in $x$ and PBC in $y$. 
   Parameters: $\lambda=0.5$, $K=0.85$, and $y$ length is $10 L_y$ with
    $k_y^{\max}=0.53$. Domains and times are (from left to right): $L_x/L_K=7$
    ($t=260$), $L_x/L_K=15$ ($t=640$),
    $L_x/L_K=20$ ($t=670$), and $L_x/L_K=30$ ($t=675$). 
(b) Distance ($\Delta L_D$) between the two locations of
    initial defect formation normalized by the domain length
    ('$\times$' symbol, left axis) and the time at which they appear
    ('$\bullet$' symbol, right axis) as a function of number of
    periods for $K=0.85$. Dashed line represents a fit
    $\Delta L_D/L_x=1 -2L_D/L_x$, where $L_D\simeq 4.5$ is the roughly constant distance of
    defect location from the boundary; note the asymptotic
    limit 1 as $L_x \to \infty$ . 
(c) Snapshots of DNS 
    of~\eqref{eq:she} at times $t=231,672,3800$ for different $K$
    values but keeping $x \in [0,20 L_K]$ fixed, respectively (from
    left to right). Initial and boundary conditions as in
    (a) with $y$ length $12 L_y$, where
    $k_y=k_y^{\max}=0.57,0.53,0.47$, respectively.
}
  \label{fig:50p}
\end{figure*}

{The time evolution of a perturbed stripe is quite distinct for 
NBC (where MM dominates) compared to PBC (where ZZ dominates)}. 
For $K$ in the ZZ unstable
  range, but close to the ZZ instability line, a random perturbation
  yields the ZZ stripes under PBC [see Fig.~\ref{fig:zz_stable}(a)],
  but no instability of $u_K$ under NBC [see
  Fig.~\ref{fig:zz_stable}(b)]. For $K$ deeper in the ZZ unstable
  range, at least on a long transient scale the behavior under PBC
  only changes qualitatively, leading to stripes that bend more
  strongly, and which on even longer time scales may or may not develop
  defects. However, under NBC we now are beyond the MM line, and the
  transient behavior is dominated by a mixed mode, as illustrated in
  Fig.~\ref{dnsf1}, and where the solution has generated defects in the bulk 
  already at $t{=}500$ in (a), and at $t{=}1500$ in (b).

The characteristics of the MM instability can be examined further 
by choosing initial perturbations in the MM direction, and 
by variation of the number of periods in $x$ (i.e., copies of $L_K$), or of the distance from the Eckhaus onset. In the following, let $k_y^{\max}$ be the extremum point in the MM dispersion relation [see Fig. \ref{fig:bb}(b)]. 
Figure \ref{fig:50p} shows DNS  with initial condition
\begin{equation}\label{eq:pert_eig}
  u(x,y)=u_K(x)+\eps\, \tilde{u}_\text{MM}(x)\cos(k_yy){\big|}_{k_y=k^{\max}_y} \,,
\end{equation}
with $\eps=0.025$ and $\|\tilde{u}_\text{MM}\|_\infty=1$, over different domains (but fixed $K$, (a) and (b)) 
and for different $K$ (but fixed domain, (c)). An increase in domain size (a) shifts the initial defect formation 
to locations near the boundary, leaving the bulk to form ZZ behavior. 
A similar behavior is also observed while keeping the periodicity ($20 L_K$ in $x$ direction) and $\lambda$ fixed and approaching the Eckhaus onset by decreasing $K$,
as shown in (c). We attribute the evolution of defects near the boundary in both cases to the competition between the ZZ and mixed modes. If 
$L_x$ is large enough, then the bulk preserves locally the phase symmetry, 
and thus is primarily subjected to the ZZ mode. 
In (b) we show the spatial spacing between the defects and the time at which the first appearance of the defects is emerged while indicating in (a) that for the given parameters the defects transiently form an interface between the straight rolls at the boundary and the ZZ rolls in bulk. Moreover, once the domain is long enough, also the time scale for the appearance of defects saturates. The same applies once the Eckhaus onset is approached since the MM is more prominent there than near the MM onset, where the defects again form in the bulk after a very long transient, as shown in (c). We have also investigated the influence of finite size effects in $y$-direction with respect to the location of defects, i.e., phase effects. Here, DNS did not show any influence (besides changes in the time scales) of the variation of domain size, or of NBC in $y$. 
\begin{figure*}[tp]
\btab{ll}{(a)&(b)\\
\ig[width=0.33\tew]{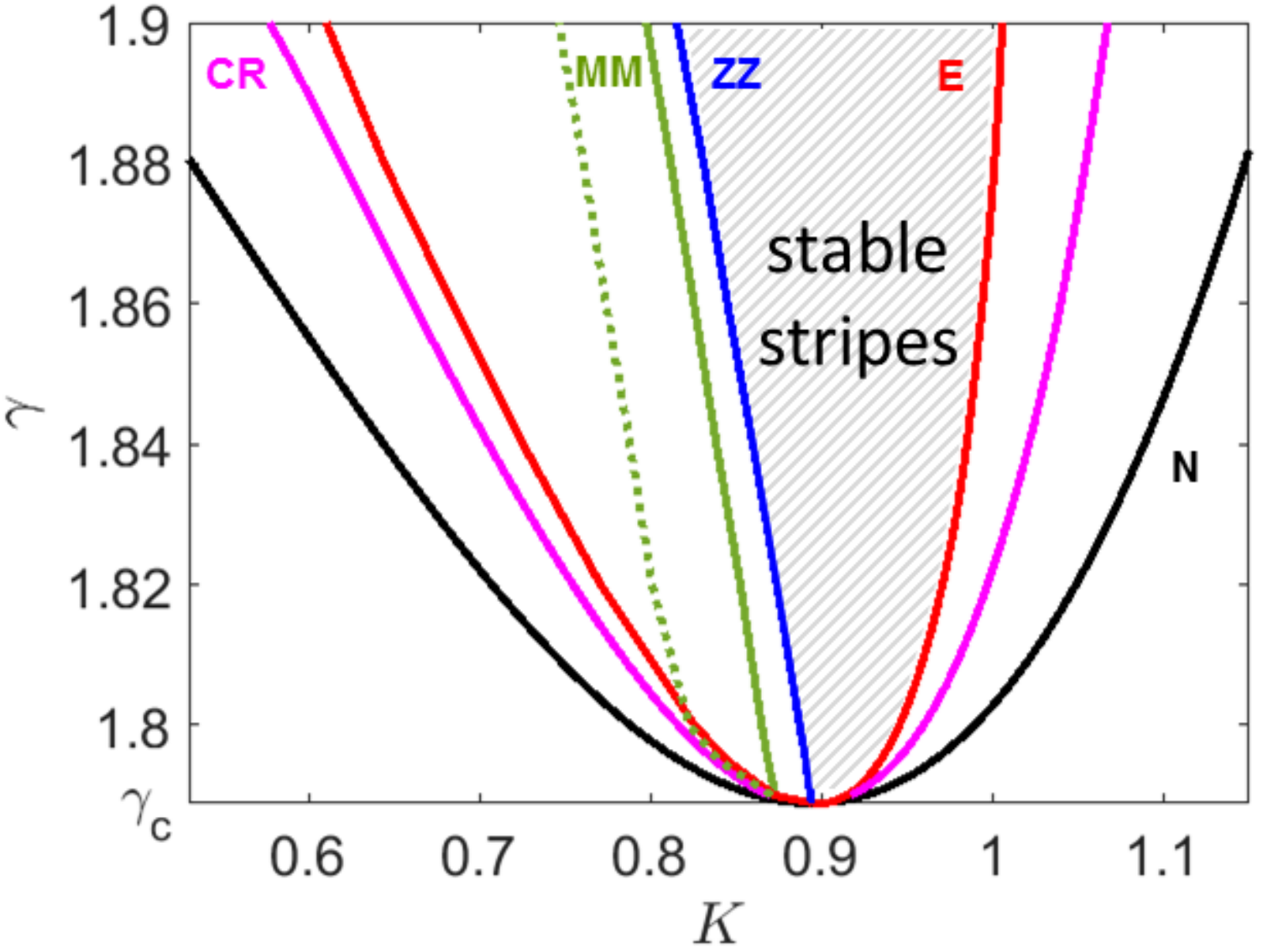}\hs{5mm}
&\raisebox{3mm}{\ig[width=0.6\tew]{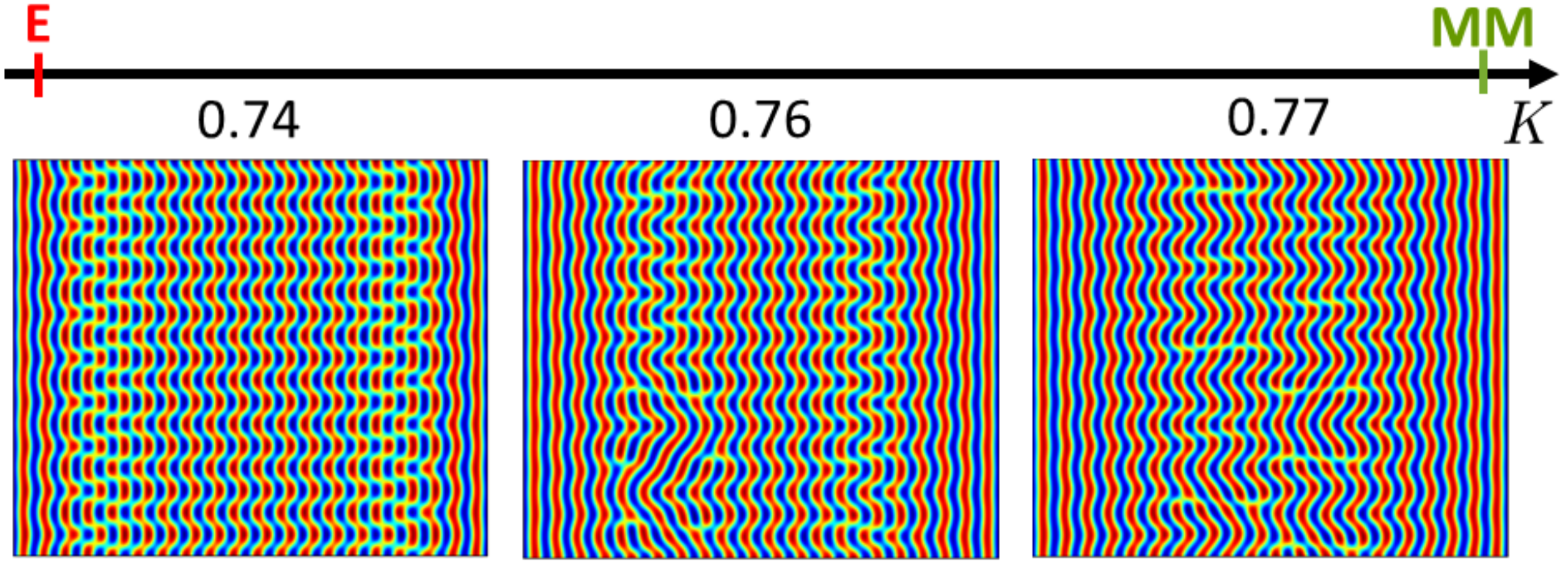}}}
  \caption{(a) Existence and stability ranges of (periodic) stripe
    solutions of the FCGL equation~\eqref{eq:cgl}, notations and numerical
    details as in Fig.~\ref{fig:bb}(a). Again, the blue ZZ line only pertains to PBC and for NBC is ``replaced'' by the MM line. (b) Snapshots showing the
    direct numerical integration of~\eqref{eq:cgl} at $\gamma=1.84$
    and times $t=580,2080,4400$ for different $K$ values, respectively
    (from left to right); colorscale represents the minimal (blue) and the maximal (red) values of $\Re A$, boundary and initial conditions analogous to Fig.~\ref{fig:50p}(c). Other parameters: $\mu=\beta=0$, $\nu=2$,
    $\alpha=0.5$, $\Omega=[0,20 L_K]\times[0,9 L_y]$, where
    $k_y=k_y^{\max}=0.42,0.39,0.37$, respectively.
  }\label{fig:cglsim}
\end{figure*}

\section{The forced complex Ginzburg-Landau equation}
To substantiate further the generality of the MM for stripe instability 
on finite domains, we next consider 
the forced complex Ginzburg-Landau (FCGL) equation, which 
is known to exhibit a finite
wavenumber instability in the 2:1 resonance case~\cite{Yochelis2002}, 
and in contrast to 
the SH equation is not a gradient system. 
It reads  
\begin{equation}\label{eq:cgl}
  \parf{A}{t}=(\mu+i\nu)A-(1+i\beta)|A|^2A+\gamma A^*+(1+i\alpha)\nabla^2A,
\end{equation}
with $A\in \mathbb{C}$ (and $A^*$ denoting the complex conjugate), 
and parameters $\mu,\nu,\beta,\alpha,\gamma \in \mathbb{R}$, 
where we shall use $\gamma$ as the instability parameter.  Although  
 \reff{eq:cgl} can describe various circumstances, such as chemical oscillations~\cite{Yochelis2002} and nonlinear optics~\cite{gomila2004stable}, here we 
consider it simply as a two--variable second order reaction--diffusion 
system with a generic behavior near the Turing onset.  
 The trivial state $A=0$ shows  instabilities 
of Hopf-Turing co--dimension 2 type. We focus here only on the 
Turing onset and steady spatially periodic solutions by keeping the Hopf mode neutral (i.e., $\mu=0$) so that oscillatory solutions have zero amplitude. In this case, the pure Turing solutions bifurcate from the onset $\gamma_c{=}\nu/\rho$
with critical wavenumber $K_c^2{=}\nu\alpha/\rho^2$,  where $\rho=\sqrt{1+\alpha^2}$ 
~\cite{yochelis2004two}.


We follow the same methodology as for the SH model and compute the 
ZZ, E, CR and MM onsets, and find that also for the FCGL equation 
the MM onset
lies to the left of the ZZ line and depends on the domain size, as
shown in Fig.~\ref{fig:cglsim}(a). Additionally, DNS 
using NBC confirms the dominance of the MM {on the left of the ZZ onset (with PBC)}, with defects being formed near the boundaries 
in $x$, as shown
in Fig.~\ref{fig:cglsim}(b).

\section{Discussion}
{We have characterized a distinct impact of domain size and boundary conditions on the instability of stripes. Using two prototypical models, the (variational) Swift-Hohenberg and the ({non--variational}) forced complex Ginzburg-Landau equations, we showed through numerical analysis the existence of a distinct secondary \textit{mixed-mode} instability in between the Eckhaus and the zigzag onsets. The instability is a direct and generic consequence of deviation from the infinite domain 
assumption (or large domain with PBC) on which the analysis is typically performed~\cite{hoyle2006pattern,nepomnyashchy2006general,meron2015nonlinear}.} This MM instability results under Neumann BC and mixes properties of the ZZ and Eckhaus instabilities, and in DNS triggers transient defects first near the domain boundaries, as shown in Fig.~\ref{fig:50p} {and Fig.~\ref{fig:cglsim}}. 
The locations where these defects form are solely related to the amplitude decay of the eigenfunction (see Fig.~\ref{fig:bb}(c)), exactly as for the Eckhaus instability albeit with a non zero $k_y$.

We believe that our insights will be valuable for understanding stripe
pattern evolution at early stages and their sensitivity to BC,
especially for systems that inherently exhibit large separation of
time scales, such as soft matter electrochemical
media~\cite{gavish2016theory,shapira2019pattern}, developmental
biology~\cite{smith2008role,marcon2012turing,xu2017turing,landge2020pattern}
and vegetation
patterns~\cite{meron2019vegetation,bastiaansen2020effect}. {Moreover,
in physicochemical systems with practical applications, small domains
are often of interest, e.g. thin layers in organic
photovoltaics~\cite{bedeloglu2010photovoltaic}, highly concentrated
electrolytes~\cite{pontoni2017self}, and superconducting quantum
interference device metamaterials~\cite{hizanidis2020pattern}. In such systems, stripe morphology may persist (shortly)
beyond the analytically expected ZZ onset. On the other hand, 
if the MM line is crossed, then stripes can be also more sensitive to perturbations: Defects may form near the boundary, yielding breakups on a relatively 
short time scale.}

\begin{acknowledgments}
	We thank Edgar Knobloch (UC Berkeley) for helpful discussions. The research was supported by the Adelis Foundation for renewable energy research.
\end{acknowledgments}
\vskip 0.2in
The data that support the findings of this study are available from the corresponding author upon reasonable request.
\vskip 0.2in
\bf{REFERENCES}

\end{document}